# Dissecting the Gender Divide: Authorship and Acknowledgment in Scientific Publications

Keigo Kusumegi[1], Daniel E. Acuña[2], Yukie Sano[3*]

[1]Department of Information Science, Cornell University, Gates Hall, 107 Hoy Road, Ithaca, 14853, NY, USA. [2]Department of Computer Science, University of Colorado Boulder, ECOT 717, 430 UCB, Boulder, 80309-0430, CO, USA. [3*]Institute of Systems and Information Engineering, University of Tsukuba, 1-1-1 Tennodai, Tsukuba, 305-8573, Ibaraki, Japan.

**Abstract**

The issue of gender bias in scientific publications has been a topic of ongoing debate. One aspect of this debate concerns whether women receive equal credit for their contributions compared to men. Conventional wisdom suggests that women are more likely to be acknowledged than listed as co-authors. In this study, we analyze data from over 20,000 authors and 60,000 acknowledged individuals across nine disciplines in open-access journals. Our results confirm persistent gender disparities: women are more frequently acknowledged than credited as co-authors, especially in roles involving investigation and analysis. To account for status and disciplinary effects, we examined collaboration pair composed of highly cited and less-cited scholars. In collaborations, highly cited scholars are more likely to be listed as an author regardless of gender. Notably, highly cited women in such pairs are even more likely to be co-authors than their men counterparts. Our findings suggest that power dynamics and perceived success heavily influence the distribution of credit in scientific publishing. These results underscore the role of status dynamics in shaping authorship and call for a more nuanced understanding of how gender, power, and recognition interact in scientific publishing. Our findings offer valuable insights for scholars, editors, and funding committed to advancing equity in science.

## 1 Introduction

The story of Rosalyn Franklin, a key contributor to the pivotal discovery of the structure of DNA, serves as a powerful reminder of gender bias in the academic landscape (Rossiter 1993). Despite her pivotal role, she was not given due credit as an author (Klug 1968). Instead, she was acknowledged in the DNA structure paper (Watson and Crick 1953). Another illustrative case is that of Mary Tsingou, who played a central role in one of the earliest computational experiments in physics in the 1950s. Although she was the one who implemented the numerical simulation that revealed unexpected nonlinear dynamics, her name was not listed as an author of the paper (Fermi, Pasta, and Ulam 1955) and mentioned only in the acknowledgments (Dauxois 2008).

While these examples occurred in the 20th century, the dynamics of gender and power continue to be highlighted even in the 21st century (Ross et al. 2022). These cases prompt us to consider how frequently such dynamics influence credit in science. Although gender disparities still exist in various aspects of academia, including citation practices, authorship patterns, academic representation, and dropout rates (Teich et al. 2022; West et al. 2013; Paul-Hus et al. 2020; Kong, Martin-Gutierrez, and Karimi 2022; Liu et al. 2023; Jadidi et al. 2018), the issue of authorship versus acknowledgment with respect to gender, though discussed (Rossiter 1993), remains poorly understood due to lack of comprehensive data. In this study, we aim to address this gap by examining a large-scale dataset on authorship and acknowledgment, through the lens of gender and power dynamics.

Gender disparity in academia has become a widely recognized and persistent topic of discussion, transcending disciplinary boundaries and geographical boundaries. A recent study shows that women are systematically credited less than men in science (Ross et al. 2022). Beyond individual variations influenced by factors such as career length, personal circumstances, or parental leave, women and minority groups often face systematic under-valuation within homophilic networks that favor men and reinforce cumulative advantages (Neuhäuser et al. 2023; Karimi et al. 2018; Teich et al. 2022; DiPrete and Eirich 2006). Given growing evidence that gender diversity enriches perspective-taking, collective intelligence, and innovation (Nielsen et al. 2017; Yang et al. 2022), prompting gender equality is imperative—not only to benefit individuals, but to accelerate scientific progress at large.

The assignment of credit including both authorship and acknowledgment is critical in shaping inequalities within the scientific community. Authorship, as the more prestigious form of credit, carries particular weight: it shapes researchers' professional profiles and influences their evaluation within institutional reward structures (Maddox 2002; Cobb and Comfort 2023). However, biased credit allocation persists, disproportionately disadvantaging early-career researchers and influencing processes such as hiring, funding, and prestigious awards. Gender disparities further compound these issues. Prior research shows that women, despite making comparable contributions, are less likely to be credited as authors or acknowledged appropriately (Ross et al. 2022). Moreover, women frequently encounter disputes over authorship order and naming, leading to frustration and discouragement when their contributions go unrecognized (Ni et al. 2021). These disparities demand urgent attention to rectify gender biases in credit assignments, particularly authorship, given its profound impact on careers achievement and access to scientific opportunities.

The concept of credit in scholarly publications has undergone a significant shift, moving from a traditional emphasis on authorship to a greater recognition of individuals' specific contributions, particularly as research teams grow larger and collaboration extends beyond geographical boundaries (Holcombe 2019; Vasilevsky et al. 2021; Fortunato et al. 2018). In 1997, Rennie et al. proposed a contribution-based system that aimed to accurately thou scholarly contributions, offering a valuable framework for scientific institutions and society to assess researchers' actual contributions rather than relying solely on authorship (Rennie, Yank, and Emanuel 1997). In 2016, PLOS introduced the Contributor Roles Taxonomy (CRediT) system, a standardized system for recognizing the diverse contributions made to scholarly publications (Sauermann and Haeussler 2017; Corrêa Jr et al. 2017). Larivière et al. analyzed CRediT data from PLOS publications between 2017 and 2018, revealing systematic associations between contributor roles, author counts, and author position (Larivière, Pontille, and Sugimoto 2021). This shift towards recognizing specific contributions has also emerged beyond academia. For example, in open-source software development, where recognizing a range of contributions offers more comprehensive picture of collaboration (Young et al. 2021). Together, these developments highlight the growing importance of accurately attributing contributions and providing the full complexity of collaborative work across domains.

With the increasing emphasis on recognizing specific scholarly contributions, concerns about gender equality in these contributions have emerged. Previous research has examined the gender differences in authorship roles and found that women are more likely to be associated with performing experiments, whereas men are more frequently credited with other roles, even after accounting for academic age (Macaluso et al. 2016). However, the issue lies in biased credit assignment: women may be acknowledged rather than listed as co-authors, despite making similar contributions. In this context, a comparative analysis of authors and acknowledgees (those acknowledged in a paper) becomes crucial for evaluating how credit is assigned and whether such practices are equitable across genders.

Research on scientific credit assignment has also explored whether power dynamics, beyond gender and race, play a role. These dynamics often manifest in institutional hierarchies, where highly-cited senior scholars exert influence over less-cited colleagues (Aiston and Jung 2016). Power imbalances are also evident in research funding, where decision-making bodies are predominantly composed of established scholars (Lauer and Roychowdhury 2021). These individuals tend to favor research topics and investigators that align with their own research interests, further skewing the credit assignment process. Thus, the interplay of power dynamics in scientific credit assignment extends beyond personal attributes like gender and race, and warrants systematic scrutiny to ensure equity and fairness in recognition of academic contributions.

In this paper, we aim to investigate the gender disparities in the attribution of scientific credits, focusing in particular on authorship and acknowledgment by analyzing the scholarly contributions of various disciplines. Building on previous work that have identified acknowledged scholars (Kusumegi and Sano 2022), we leverage this data to match authors and acknowledgees and compare their respective contributions. To investigate this phenomenon, we adopt two primary analytical strategies.

First, we explore whether gender differences can explain why some scholars receive co-authorship roles, while others are relegated to acknowledgments across different publications. Simultaneously, we examine how power dynamics may influence these credit assignments. Second, we perform an analysis that accounts for various individual-level factors, focusing specifically on disparities within research teams where the same scholars have appeared either as co-authors or acknowledgees in different publications. We further analyze the implications of these credit disparities in terms of citation outcomes, assessing the extent to which misalignment between credit received and contributions made affects scholarly visibility.

By understanding the gender imbalance in acknowledgments, we aspire to promote equitable scholarly assessments. Our findings offer a more comprehensive perspective that goes beyond the conventional focus on authorship alone. By incorporating power dynamics into the analysis, we demonstrate how credit allocation is shaped not only by contributions but also by perceived status and predetermined roles within the academic sphere.

In light of the complexities and ambiguities surrounding gender bias in authorship and acknowledgment within academic research, this study aims to answer the following research questions:

1. Are women more likely to be acknowledged rather than listed as co-authors across various contribution roles?

2. How is academic status, as measured by citation count, related to whether a contributor is listed as an author or acknowledged?

By answering these questions, we aim to contribute to the ongoing discussions about gender bias in scientific research.

## 2 Materials and methods

### 2.1 Authorship and acknowledgment data

We extracted authorship and acknowledgment data from seven open-access journals: PLOS Biology, PLOS Computational Biology, PLOS Genetics, PLOS Medicine, PLOS Pathogens, PLOS Neglected Tropical Diseases, and PLOS ONE. These journals were selected due to their accessibility and the availability of acknowledgment statements in their publications. Specifically, we focus on papers published between 2016 and 2021 as PLOS introduced CRediT (Contributor Roles Taxonomy) system in 2016 (https://theplosblog.plos.org/2016/07/author-credit-plos-and-credit-update/). CrediT is a system that standardizes the way in which contributors to scholarly works are acknowledged and credited. It aims to increase transparency regarding individual roles in research and promote fairness in credit allocation.

For acknowledgment data, we utilize the dataset of identified acknowledged scholars (Kusumegi and Sano 2022). This dataset, based on PLOS journals, provides DOIs and the names and unique identifiers of acknowledged scholars, linked to Microsoft Academic Graph (MAG) IDs (Sinha et al. 2015). Here we utilized the snapshot of MAG data until 2021-12-31. Today, a compatible alternative is OpenAlex (https://openalex.org/), which continues the structure and identifier system of MAG and is publicly available for ongoing research. By combining each scholar's name with their corresponding acknowledgment statement, we are able to trace who is acknowledged, in what context, and in which publication.

We then performed gender detection for all authors and acknowledgees. We first removed the scholars with only the first letter of their given name. Gender was then predicted using by the Gender API (https://gender-api.com/), a service that infers a person's gender based on their first name. This service has shown to have good performance compared with other gender-detecting services (Sebo 2021).

In this study, we also refered the Field-of-Study taxonomy in MAG, which is organized hierarchically, from broad disciplines (level 0) to highly specific subfields (level 5). For our analysis, we used the level 0 classification to represent academic disciplines, which includes broad categories such as Medicine and Biology.

## 2.2 Contributor role

We defined the roles of contributors (i.e., authors and acknowledgees) according to the following rules.

### 2.2.1 Authorship contribution role

The role of authors in scholarly contributions is important as it offers credit and has academic and financial implications. The Contributor Roles Taxonomy (CRediT) provides transparency in authorship by individual contributions to published work, thereby improving systems of attribution, recognition, and accountability. Since around 2016, PLOS journals have adapted author roles via the CRediT taxonomy instead of the traditional contribution statements.

To facilitate comparison with acknowledgment-based contribution roles, we map the CRediT taxonomy onto a more abstract classification scheme (Table 1). Compared to roles found in acknowledgments, the authorship roles include both additional categories and some diminished (less represented or collapsed) categories. Certain authorship-specific roles, such as Funding, Administration, and Conceptualization, are not typically included among acknowledgment roles. According to our taxonomy, for example, if author A's contributions in the CRediT taxonomy include "writing – original draft preparation" and "data curation," then A's authorship role is classified as "Writing" and "Investigation and Analysis".

Table 1. Taxonomy for authorship contribution.

| Category | CRediT |
| --- | --- |
| Investigation and Analysis | co-investigator, data curation<br>formal analysis, investigation<br>principal investigators, research assistants<br>software, methodology<br>validation, visualization |
| Material and Resources | resources |
| Writing | writing – original draft preparation<br>writing – review & editing |
| Funding | funding acquisition |
| Administration | project administration, supervision, |
| Conceptualization | conceptualization |

### 2.2.2 Acknowledgment contribution role

Acknowledgments in research publications indicate meaningful contributions to scientific work. However, unlike authorship credit where CRediT provides a standardized taxonomy, no universal framework yet exists for categorizing contributions made through acknowledgments. Acknowledgments serve as a space to recognize contributions that are essential but not sufficient for authorship, offering insight into hidden yet valuable aspects of research progress (Xie and Zhang 2022). This lack of standardization makes it challenging to establish consistent taxonomy for acknowledged contributions. Generally, acknowledgment frequently mention support such as technical support, financial support, and peer communication (Tiew and Sen 2002; Nimale, Khaparde, and Alhamdi 2015; Paul-Hus, Díaz-Faes, et al. 2017).

Contributions in acknowledgments can often be identified by the specific words used in each context. Paul-Hus et al. developed a codebook composed of 13 categories of acknowledgment and investigated the proportion of the frequently appeared nouns in each category (Paul-Hus and Desrochers 2019). For example, the word "analysis" appears at 47% in "Investigation and Analysis" and 32% under "Disclaimer." Building on their framework, we constructed our own taxonomy of acknowledgment contributions based on both this categorization and frequency of word usage.

In particular, we consider words that appeared in a single category with more than 60% frequency as reported in Paul-Hus et al. (Paul-Hus and Desrochers 2019), to be indicative of the corresponding contribution type. For example, the word "discussion" appears in the context of "Peer communication," 98% of the time. Therefore, a sentence like "We appreciate the discussion with $A$" interpreted as indicating $A$'s contribution as a role of "Peer communication." For words that do not show a strong association with a single category, we manually examined their usage in our dataset through random sampling. Details of this manual survey are provided in the Appendix A.

Note that the frequency of certain words in our dataset varies from the original statistics (Paul-Hus and Desrochers 2019). One possible reason for these differences is the nature of the data sources. While Paul-Hus et al. (Paul-Hus and Desrochers 2019) used the data from Web of Science, which focuses on funded research, our dataset includes papers both with and without funding. Another possible reason can be that our dataset only considers interpersonal relationships. Because our acknowledgment dataset (Kusumegi and Sano 2022) identifies individuals by name, papers containing only organizational or non-personal acknowledgments were excluded. For example, this may explain the under-representation of the "Disclaimer" category, which typically reflects authors' statements rather than interpersonal contributions.

Based on these considerations, we constructed a simplified taxonomy consisting of four categories used for analysis: "Investigation and Analysis," "Material and Resources," "Writing," and "Peer Communication," along with representative keywords for each category (Table 2). Although "Disclaimer" was initially identified as a potential category, it was excluded from the final taxonomy used in the analysis due to its lack of interpersonal contribution.

Table 2. Taxonomy for acknowledgment contribution.

| Category | Keywords |
|---|---|
| Investigation and Analysis | assistance, experiment<br>help, measurement<br>analysis, collection<br>design, interpretation<br>code, data<br>work, preparation |
| Material and Resources | access, data |
| Writing | writing |
| Peer communication | discussion, review |

Once the taxonomy of acknowledgment contributions is established using noun-based keywords, we can identify who is acknowledged for which type of contribution by extracting nouns from acknowledgment statements. For this process, we used the Python package nltk (https://www.nltk.org/). First, each acknowledgment statement is segmented into sentences, and nouns are extracted from each sentence using lemmatization. If a noun that matches a keyword from our acknowledgment taxonomy (Table 2) appears in a sentence along with a personal name, we infer that the individual was acknowledged for the corresponding contribution type. For example, in a sentence such as "We thank *A* for helpful discussion", we consider that *A* is acknowledged as a role of "Peer Communication" by the keywords of "discussion." Multiple individuals and contribution roles can be identified within a single sentence.

## 2.3 Authorship Rate (*AR*)

We examine gender disparities in authorship and acknowledgment with a focus on specific contribution roles. This analysis is conducted using the Authorship Rate ($AR$), defined for gender $i$ and contribution role $k$ as:

$$AR_i^k = \frac{\text{Number of authors of gender } i \text{ in role } k}{\text{Number of contributors of gender } i \text{ in role } k}$$

where "contributors" refers to individuals who are either authors or acknowledgees. Hereafter, we omit the notation $k$ and $i$ for simplicity. To clarify the gender difference in $AR$, we calculate relative difference in average $AR$ between women and men, calculated as $\frac{E(AR_{\text{women}})-E(AR_{\text{men}})}{E(AR_{\text{men}})}$, where $E(x)$ denotes the mean of $x$.

Using the $AR$ measure, we evaluate gender differences in contribution roles at both the paper level and the collaboration level (Fig. 1). The paper-level analysis focuses on gender balance among contributors within individual papers, while the collaboration-level analysis captures gender dynamics in interpersonal collaborations between men and women. The fundamental concept behind this evaluation is that when men and women contribute equally to the same specific role, they should be equally recognized as authors. This approach is particularly meaningful for contribution roles that appear in both authorship and acknowledgment, namely, Investigation and Analysis (I&A), Materials and Resources (M&R), and Writing.

In the paper-level analysis (Fig. 1(a)), the $AR$ can be interpreted as the fraction of individuals recognized as authors out of the total number of individuals identified either as authors or in acknowledgments. To compare the average $AR$ between men and women for each contribution role, we conducted independent two-sample $t$-test. This analysis includes all authors and acknowledgees, regardless of whether they are identified as scholars—i.e., even individuals without academic publication records in our dataset are counted.

In the collaboration-level analysis, we calculate the $AR$ for each man-woman pair of collaborators (Fig. 1(b)). For example, if Alice and Bob collaborated on $n$ projects with identical contribution roles, and Alice was listed as an author $j(\leq n)$ times while Bob was listed $k(\leq n)$ times, their $AR$ scores would be $j/n$ and $k/n$, respectively. This analysis focuses on gender balance in one-to-one collaborations, rather than in larger teams or group-level dynamics. A paired $t$-test was used to compare the average $AR$ of men and women for each contribution role. Since scholar IDs are required to track individual publication records, acknowledgees who do not have a corresponding ID in the MAG were excluded from this analysis.

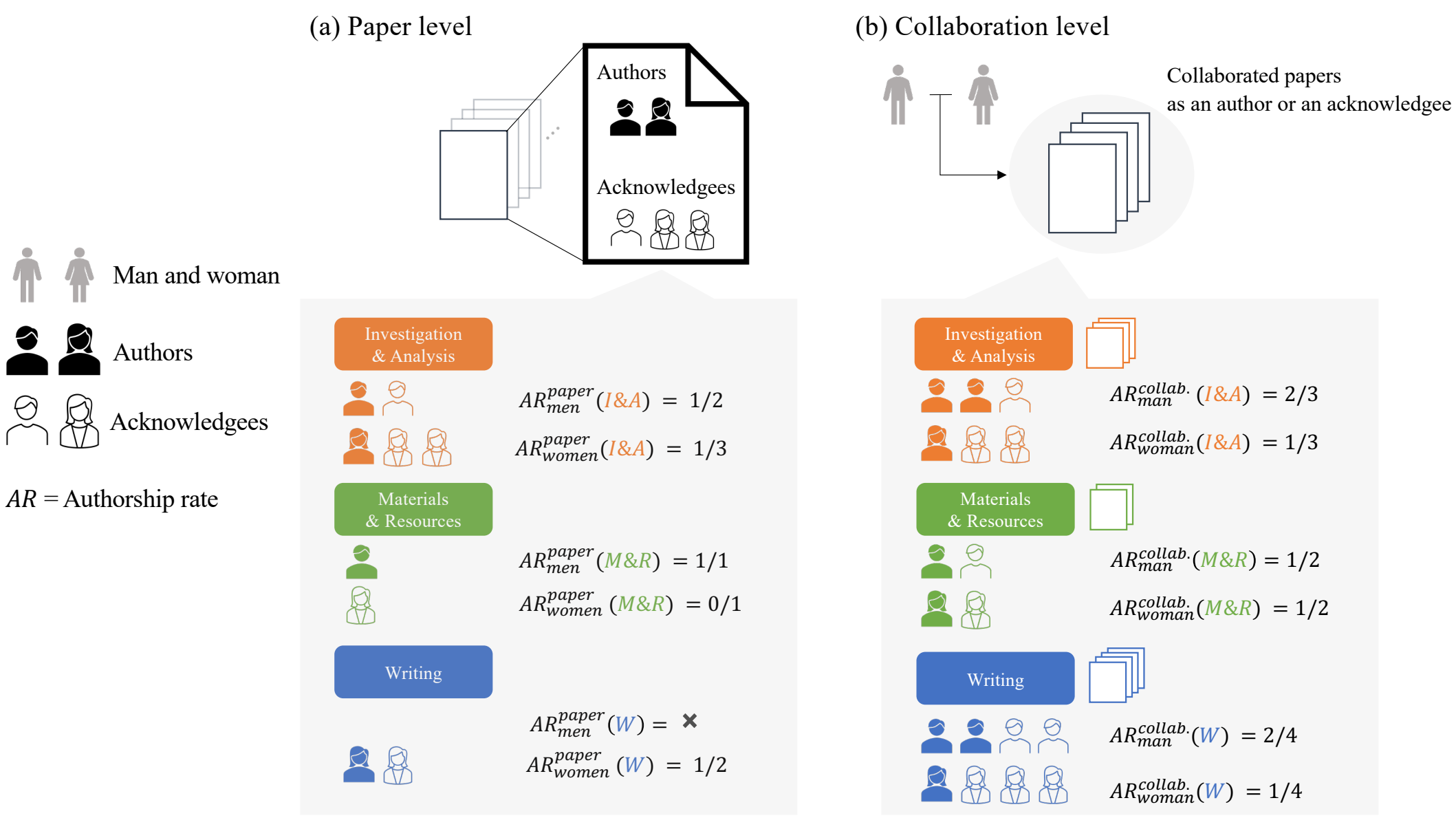

Fig.1 The analysis process from extraction of contribution to test statistics. (a) Paper-level analysis: Gender differences are evaluated within each paper. Authors and acknowledgees (i.e., individuals named in acknowledgments) are extracted along with their contribution roles. If both men and women contributions are involved in a given role, the Authorship Rate ($AR$) is calculated separately for each gender. For example, in the role of I&A, if one man author and one man acknowledgee contributed, the man $AR$ is $1/2$. If one woman author and two women acknowledgees contributed, the women $AR$ is $1/3$. (b) Collaboration-level analysis: Gender differences are evaluated in man–woman collaborator pairs, regardless of whether credit was given as authorship or acknowledgment. For each pair, $AR$ scores are computed across all papers in which both individuals contributed to a given role. For example, if a woman–man pair coauthored three papers, and the man contributed to I&A as an author in two of them while the woman did so in one, their $AR$ scores for I&A would be $2/3$ and $1/3$, respectively. A pair can share multiple roles (e.g., I&A and Writing) within the same paper.

## 2.4 Scholarly status

For collaboration-level, we conducted an analysis to examine the relationship between status differences, as measured by citation counts. Specifically, we identified scholars whose citation counts fell within the top 10% (highly cited) and the bottom 10% (less cited) of the distribution and compared gender differences in $AR$ based on their collaboration patterns. The sample included as follows: 10,586 highly cited men (high-men), 2,797 highly cited women (high-women), 6,893 less cited men (less-men), and 7,152 less cited women (less women). Details of the categorization by gender are provided in the Appendix B.

We then classified collaborations into four pair types for further analysis: (1) high-man and high-woman, (2) high-man and less-woman, (3) less-man and high-woman, and (4) less-man and less-woman. Note that this analysis only includes scholars who appeared both as authors and as acknowledgees at some point in the dataset, ensuring that their contributions can be compared across credit types. Furthermore, because we selected the top and bottom 10% of citation counts independently for men and women, the absolute citation thresholds—and thus the status levels they represent—differ between genders. That is, a "high-women" in our sample may have fewer citations than a "high-men", simply due to overall disparities in citation distributions by gender.

## 3 Results

### 3.1 Basic statistics

First, we display basic statistics about the number of authors and acknowledgments in our study. Our data consist of 19,885 papers, which included 122,072 authors and 63,343 individuals who were thanked or acknowledged. Among those acknowledged, 18,754 (29.6%) were identified as scholars, while the remaining 44,589 (70.4%) were not. Figure 2 (a) shows the average number of acknowledged individuals in relation to the number of authors per paper. Single-authored papers were rare, while most papers had a moderate number of authors, typically between two and ten. Interestingly, we observed a U-shaped trend in the average number of acknowledged individuals.

As the number of authors increases, the number of acknowledged individuals initially declines. This pattern consistent with prior research by Paul-Hus et al., which showed a decreasing or flat relationship up to around nine authors in disciplines such as Biology, Social Sciences, and Professional Fields (Paul-Hus, Mongeon, et al. 2017). However, beyond this range (around ten authors or more), the average number of acknowledgees begins to rise again, suggesting a new mechanism that may not have been previously captured. This upward trend in larger author teams may reflect the unique credit allocation patterns in "big science," where a broader range of contributors are involved and formally acknowledged. In other words, while the total number of remains relatively stable in smaller teams, larger-scale collaborations may operate under distinct authorship and acknowledgment norms.

Figure 2 (b) shows the men and women counts across nine disciplines, each represented by more than 200 papers. In our dataset, Medicine and Biology are the most prominent fields. In all discipline except Psychology, the number of men contributors exceeds that of women contributors.

Figure 2 (c) illustrates the breakdown of contribution types in authorship and acknowledgment. Based on our own classification of contribution roles, we categorized author contributions into six types: Investigation and Analysis (I&A), Materials and Resources (M&R), Writing, Funding, Administration, and Conceptualization. We also sorted acknowledgees' contributions into four types: I&A, M&R, Writing, and Peer Communication. The proportions of men and women contributors were similar in both authorship and acknowledgment, as confirmed by chi-square tests ($\chi^2(5) = 0.0160, p = 0.99$ for authorship; $\chi^2(3) = 0.006, p = 0.99$ for acknowledgment). Authorship roles were dominated by I&A and Writing, which together accounted for 60% of all contributions, while Funding being the smallest contribution. In contrast, I&A was the predominant contribution type among acknowledgments, comprising over 73% for men and 80% for women. Writing appeared in less than 1% of acknowledgment roles for both genders.

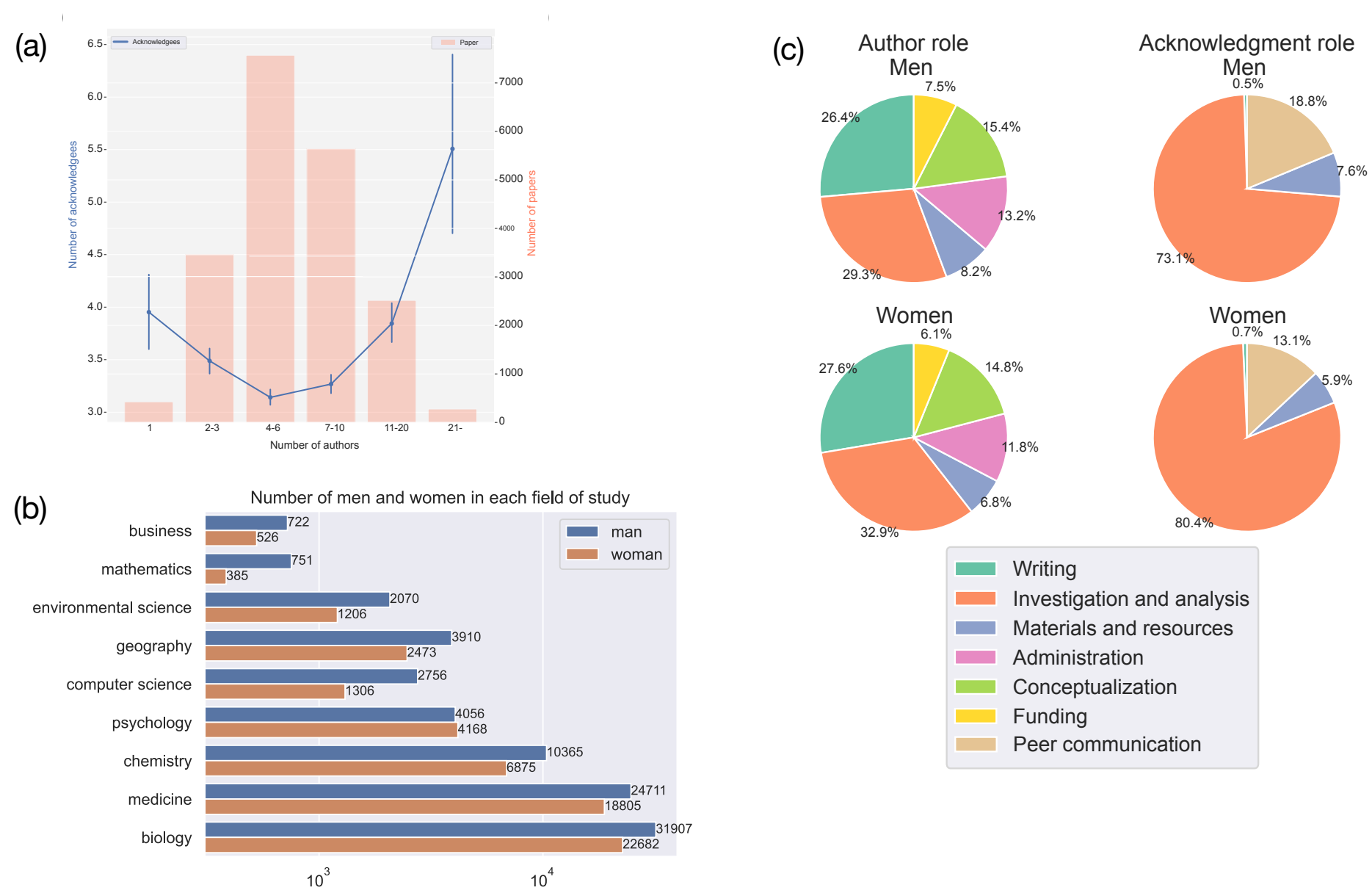

Fig.2 Descriptive statistics. (a) Average number of acknowledgees by the number of authors and total paper volume. All acknowledged names are counted, regardless of whether the individuals are identified as scholars. Error bars represent confidence intervals. (b) Number of men and women contributors across disciplines. (c) Proportion of contribution roles in authorship and acknowledgment.

## 3.2 Men are more likely to become authors

Figure 3 presents gender differences in authorship rates across two levels of analysis: (a) paper-level and (b) collaboration-level. The results show more pronounced disparities at the paper level, particularly in the roles of Investigation & Analysis (I&A) and Writing, where men exhibit significantly higher authorship rates than women. For paper-level analysis, in the I&A role, approximately 70% of men were credited as authors, compared to only 65% of women ($t(18972) = 12.46, p = 0.001$). Despite the gender gap being smaller for Writing role, it remains statistically significant ($t(15133) = 3.05, p = 0.002$).

Similar trends are also observed at the collaboration level, where the gender difference in I&A authorship rates is narrower—88% for men versus 86% for women—but still statistically significant ($t(259{,}652) = 23.49, p < 0.001$). These findings suggest that, even when women contribute through I&A or Writing, they are less likely to receive authorship credit than their men counterparts.

Conversely, for Materials & Resources (M&R), no significant gender difference were observed at either the paper level ($t(4855) = -0.58, p = 0.56$) or the collaboration level ($t(23942) = -1.67, p = 0.10$). These results indicates that men and women receive similar recognition for their contributions in M&R.

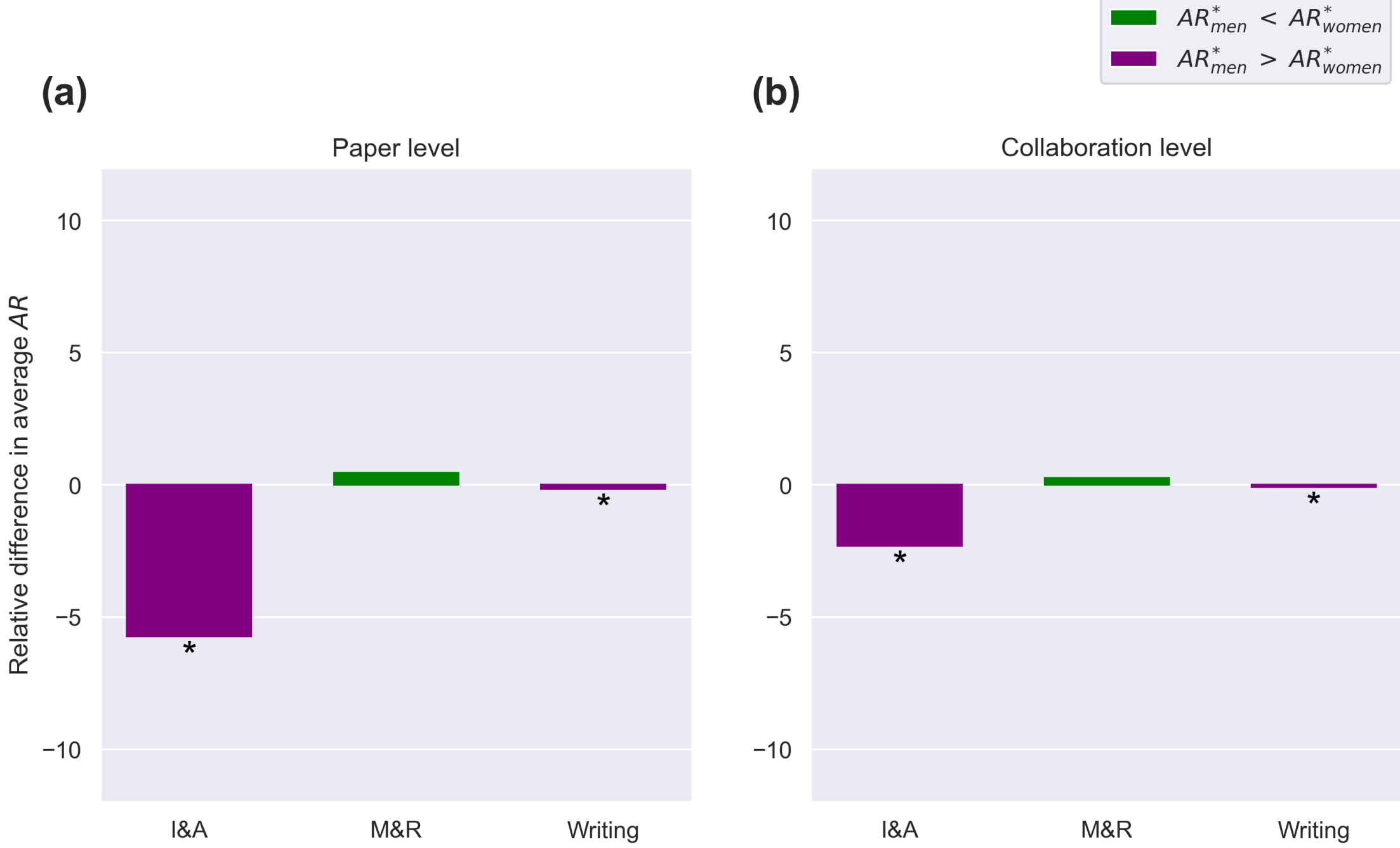

Fig.3 Gender differences in authorship rate ($AR$) by contribution role for (a) paper-level and (b) collaboration-level. Bars indicate the relative difference in average $AR$ between women and men as defined in Sec. 2.3. Purple bars indicate cases where $AR_{\text{women}} < AR_{\text{men}}$, and green bars indicate the opposite. Asterisks (*) denote statistically significant differences ($p \leq 0.05$).

A similar trend was observed across the nine disciplines we compared (Appendix C). The disparity was particularly pronounced in Medicine, while Environmental Science appeared to be the most equitable. However, since the number of publications and disciplinary cultures vary significantly, we do not explore these differences in detail here.

## 3.3 Higher status, higher authorship rate

What explains the observed gender disparities in authorship across contribution roles? Although women are less likely to be credited as authors, this bias may be influenced by other factors such as cumulative advantage or status differentials among collaborators. For instance, junior and senior scholars may have differing perceptions of what constitutes a valuable contribution, potentially leading to disagreements or undervaluation (Sauermann and Haeussler 2017). To investigate this, we examined the possibility that status asymmetries, such as collaborations between Ph.D. students and senior faculty, may underlie the patterns observed in authorship attribution. Here we conducted an analysis to examine the relationship between citation-based status differences and disparities in $AR$.

Figure 4 shows gender disparities measured by the $AR$ score across three collaboration roles. In the M&R (Materials and Resources) role, gender differences were generally small, similar to what we observed in the previous Sec. 3.2, with a significant disparity observed only in the collaboration between high-man and high-woman pairs. For the Writing role, two patterns— high-man with less-woman and less-man with less-woman—showed significant differences, with men exhibiting higher $AR$ scores in both cases.

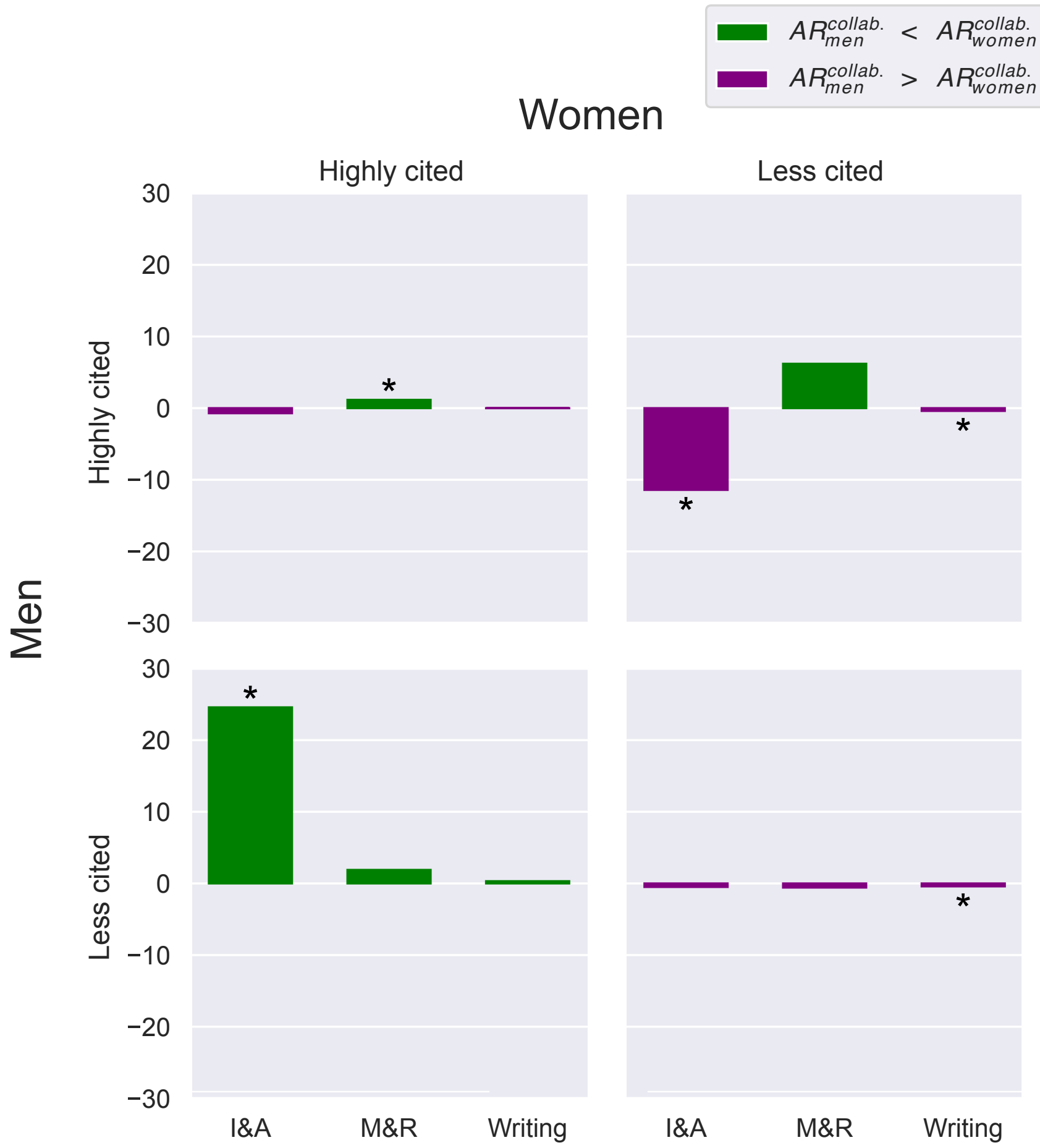

Fig. 4 Gender differences in authorship rate by citation-based collaboration patterns. The bars indicate that the relative difference in authorship rate ($AR$) between women and men, calculated using the same method as in Fig. 3. Asterisks (*) denote statistically significant differences ($p \leq 0.05$).

Substantial disparities were found in the I&A (Investigation and Analysis) role. In particular, highly cited researchers tended to receive authorship more frequently, regardless of gender. Interestingly, when a highly cited woman collaborated with a less cited man, the woman was more likely to be credited as an author. This suggests that power dynamics based on academic status or perceived success can override, or even reverse, gender-based disparities. These findings indicate that inequalities in credit assignment may be driven more by differences in scholarly standing than by gender alone.

## 4. Discussion

Our findings primarily highlight two key points: 1) Men are generally more likely to receive authorship credit, specifically in the roles of Investigation & Analysis (I&A) and Writing, with the disparity being particularly pronounced in I&A. 2) Regardless of gender, when scholars with different levels of status collaborate, those with higher citation counts are more likely to have higher authorship rate in I&R role. This pattern is evident in our dataset: across all disciplines (Appendix. C), the number of high-cited men exceeds that of women.

As we strive to address the gender disparity in authorship rates, a key question arises: "Is this inequality rooted in gender itself, or by differences in citation counts?" The answer is not straightforward. Because citation metrics, by their cumulative nature, may embed structural biases (Neuhäuser et al. 2023; Karimi et al. 2018). In particular, mechanisms such as preferential attachment (Barabási and Albert 1999) can amplify early advantages and reinforce existing inequalities over time. It requires a deep understanding of the mechanisms involved, including how credit is perceived and the underlying social and structural dynamics that shape gender differences in academic recognition. Thus, while citation counts serve as a useful indicator of academic status, they should be interpreted with an awareness of their potential limitations.

Additionally, although our highly/less citation-based classification does not perfectly align with seniority, the clear association between citation count and authorship rates $AR$ implies that social hierarchy plays a substantial role in shaping recognition. Moreover, such hierarchy is embedded within the structure of scholarly networks. Academic collaborations and citation networks often follow long-tail degree distributions (Price 1965; Newman 2001; Sheridan and Onodera 2018), and are shaped by homophily and cumulative advantage mechanisms (Merton 1968; Karimi et al. 2018). Given the historically slower entry of women into academia (Kong, Martin-Gutierrez, and Karimi 2022) and their continued minority status in many fields, women are less represented among highly cited scholars. Thus, the gender disparities we observe may partially reflect systemic bias inherent in accumulation-driven systems.

In light of these findings, efforts to ensure fair credit allocation must attend not only to overt gender bias but also to structural and hierarchical inequities. Particularly in interdisciplinary and cross-status collaborations, a nuanced understanding of how credit is assigned is essential for promoting equity in scientific recognition.

Our research is not without limitations. First, our collaboration level analysis relies on a dataset of acknowledged scholars (Kusumegi and Sano 2022), which only includes individuals with existing authorship or citation relationships. This may exclude "eternally acknowledged" individuals–those who consistently appear in acknowledgments but never as authors (Ross et al. 2022). Second, we focused on PLOS journals published after 2016 due to the adoption of the CRediT system, limiting the generalization of our findings beyond open-access venues. Third, our contribution taxonomy for acknowledgments was constructed using keyword-based classification. While CRediT roles offer a standardized framework for authorship, acknowledgment lacks an equivalent system, making role alignment challenging. Lastly, in this study, we focused on the extremes by selecting the top and bottom 10% of scholars by citation count within each discipline to identify clear patterns. However, the nature of gender disparities among the remaining 80% of scholars–the majority of the academic population–remains unclear.

## 5. Conclusions

As research becomes increasingly collaborative and interdisciplinary, the fair attribution of scholarly credit is of growing importance. This study offers an initial step toward understanding gender imbalance in credit assignment by analyzing both authorship and acknowledgment. Future work should expand beyond binary gender categories, integrate broader data sources, and further examine how network structures and institutional practices affect credit equity in science.

**Funding.** This work was supported by the Japan Society for the Promotion of Science (JSPS) KAKENHI Grant Number 23K28192 (YS).

**Declarations.** The authors have no competing interests to declare relevant to this article's content.

## Appendix A: Taxonomy of acknowledgment contribution role

When building a taxonomy of acknowledgment contribution role, we relied on the codebook based on the noun frequency (Paul-Hus and Desrochers 2019), but some words have multi meanings. In the following, we illustrate how we categorize those multi-meaning words.

A "work" is labeled as "Investigation and Analysis" unless the finance-related word is used together, such as 'foundation,' 'funded,' 'funding,' and 'grant.'

An "analysis" has multiple possibilities among "Investigation and Analysis" (47%) and "Disclaimer" (32%) according to the previous work (Paul-Hus and Desrochers 2019), but manually checking the 100 random-sampled data, the "Disclaimer" sentences are limited, and we consider "analysis" for "Investigation and Analysis."

A "preparation" has multiple possibilities among "Investigation and Analysis" (21%), "Financial disclosure" (17%), and "Disclaimer" (35%) according to the previous work (Paul-Hus and Desrochers 2019), but manually checking the 100 random-sampled data, the "Financial disclosure" and "Disclaimer" sentences are limited, and we consider "preparation" for "Investigation and Analysis."

A "data" has multiple possibilities among "Investigation and Analysis" (29%), "Materials and Resources" (40%), and "Disclaimer" (24%) according to the previous work (Paul-Hus and Desrochers 2019), but manually checking the 100 random-sampled data, the "Disclaimer" sentences are limited, and we consider "data" for "Investigation and Analysis, " or "Materials and Resources. " Here, "data" is labeled as "Materials and Resources," only if this word is used with at least one of the following keywords: 'providing,' 'provide,' 'provided,' and 'database.'

A "review" has multiple possibilities among "Peer Communication" (58%) and "Disclaimer" (25%) according to the previous work (Paul-Hus and Desrochers 2019), but manually checking the 100 random-sampled data, the "Disclaimer" sentences are limited, and we consider "review" for "Peer Communication."

A “collection” has multiple possibilities among “Investigation and Analysis” (54%) and “Disclaimer” (38%) according to the previous work (Paul-Hus and Desrochers 2019), but manually checking the 100 random-sampled data, the “Disclaimer” sentences are limited, and we consider “collection” for “Investigation and Analysis.”

A “writing” has multiple possibilities among “Writing” (16%), “Financial disclosure” (20%), “Peer Communication” (18%), and “Disclaimer” (36%) according to the previous work (Paul-Hus and Desrochers 2019), but manually checking the 100 random-sampled data, the “Financial disclosure, ” “Peer Communication, ” and “Disclaimer” sentences are limited, and we consider “writing” for “Writing. ”

A “design” has multiple possibilities among “Investigation and Analysis” (22%) and “Disclaimer” (55%) according to the previous work (Paul-Hus and Desrochers 2019), but manually checking the 100 random-sampled data, the “Disclaimer” sentences are limited, and we consider “design” for “Investigation and Analysis.”

An “interpretation” has multiple possibilities among “Investigation and Analysis” (25%) and “Disclaimer” (61%) according to the previous work (Paul-Hus and Desrochers 2019), but manually checking the 100 random-sampled data, the “Disclaimer” sentences are limited, and we consider “interpretation” for “Investigation and Analysis.”

A “code” has multiple possibilities among “Investigation and Analysis” (45%) and “Financial disclosure” (53%) according to the previous work (Paul-Hus and Desrochers 2019), but manually checking the 100 random-sampled data, the “Financial disclosure” sentences are limited, and we consider “interpretation” for “Investigation and Analysis.”

## Appendix B: Scholar’s status distribution

When grouping scholars into top 10%, bottom 10%, and the rest as middle by citation count, the numbers of them by gender and discipline are illustrated in Fig. B1.

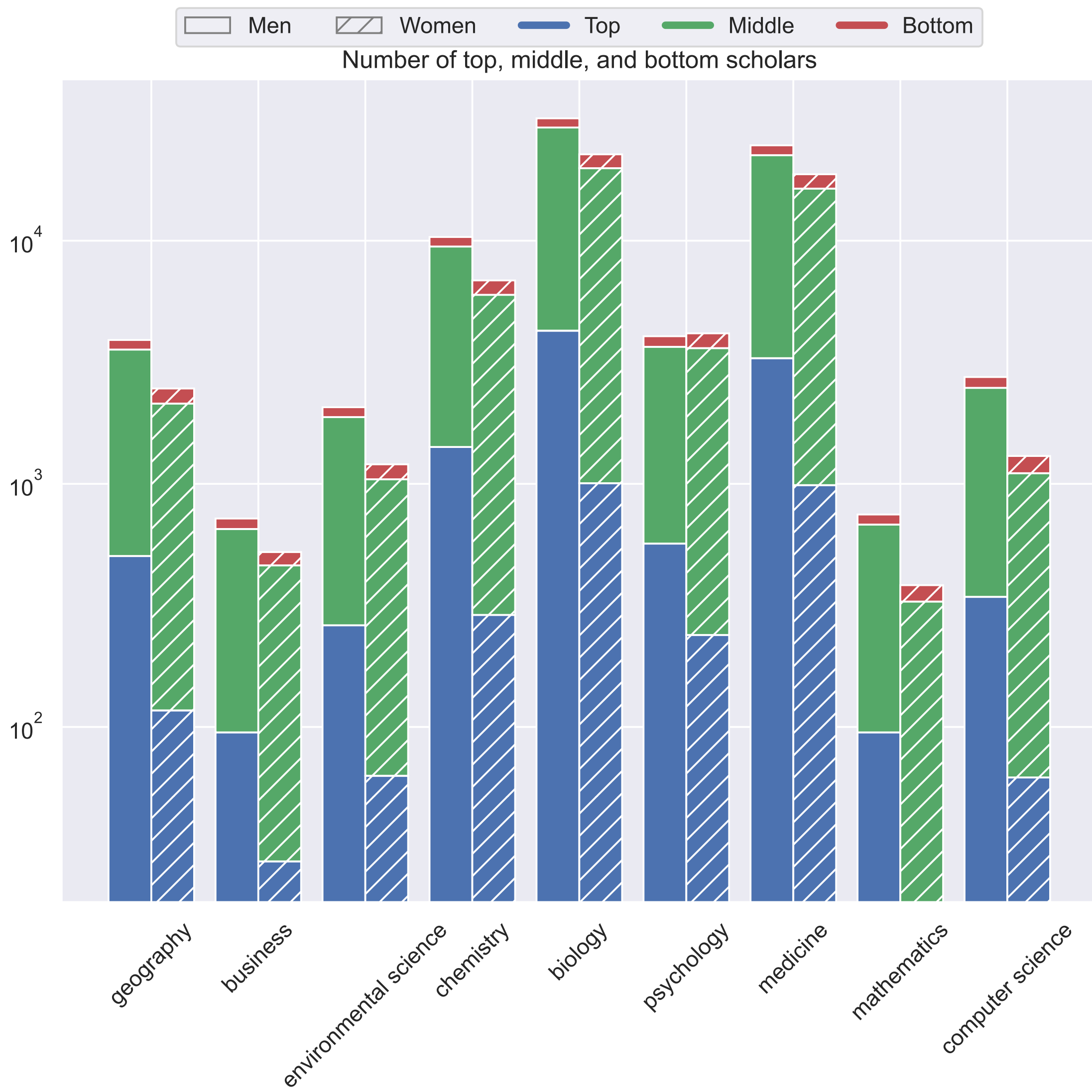

Fig. B1 Number of highly, middle, and less cited scholars by gender and discipline.

## Appendix C: Disciplinary differences in contributor roles

Upon further breaking down the results by discipline, we observed differing outcomes across various fields of study. At the paper level (Fig. C1), I&A emerges as the most gender-biased role, particularly in Medicine, Chemistry, Mathematics, Computer Science, Biology, and Psychology. Notably, Medicine, Mathematics, and Computer Science exhibit the largest gender gaps, at around 10%. The role of Writing presents a significant gender disparity only in Medicine.

The likelihood of a scholar being granted authorship for a given contribution varies greatly across disciplines. For I&A, about 70% of scholars in Medicine and Chemistry are likely to be authors, compared to about 60% in Geography, Psychology, and Environmental Science. Even larger disparities are observed in the M&R role; close to 90% of both men and women receive authorship in Medicine, Chemistry, and Biology, whereas the proportion drops to 60% in Business. It is also noteworthy that nearly all individuals across all disciplines are acknowledged as authors for their contribution to Writing.

On the collaboration level (Fig. C2), we noticed an equal or even greater gender disparities across most disciplines, with the exceptions of Psychology and Environmental Science. Interestingly, gender gaps in contribution roles often emerged at the collaboration level, even when no such disparities were observed at the paper level.

Biology, Computer Science, and Medicine appear as the most gender-imbalanced fields at the collaboration level, where all three roles–Writing, I&A, and M&R–show statistically significant disparities. Interestingly, in Computer Science and Biology, women contributing through M&R exhibit markedly higher authorship rates. Despite the general trend of men being more likely to receive authorship across most roles and disciplines, these findings suggest that women contributing via M&R in these two fields are more likely to be credited as authors than their male counterparts. Environmental Science stands out as the only discipline that exhibits no significant gender bias at both the paper and collaboration levels.

Beyond gender disparities, $AR$ scores at the collaboration level are consistently higher than those at the paper level. For instance, in Psychology, Environmental Science, Business, and Geography, the $AR$ scores in I&A and M&R at the collaboration level exceed their counterparts at the paper level by over 10%. This pattern suggests that individuals who participate in multiple collaborations, where credit is given through both authorship and acknowledgment, are more likely to receive authorship credit overall compared to those involved in fewer or single contributions.

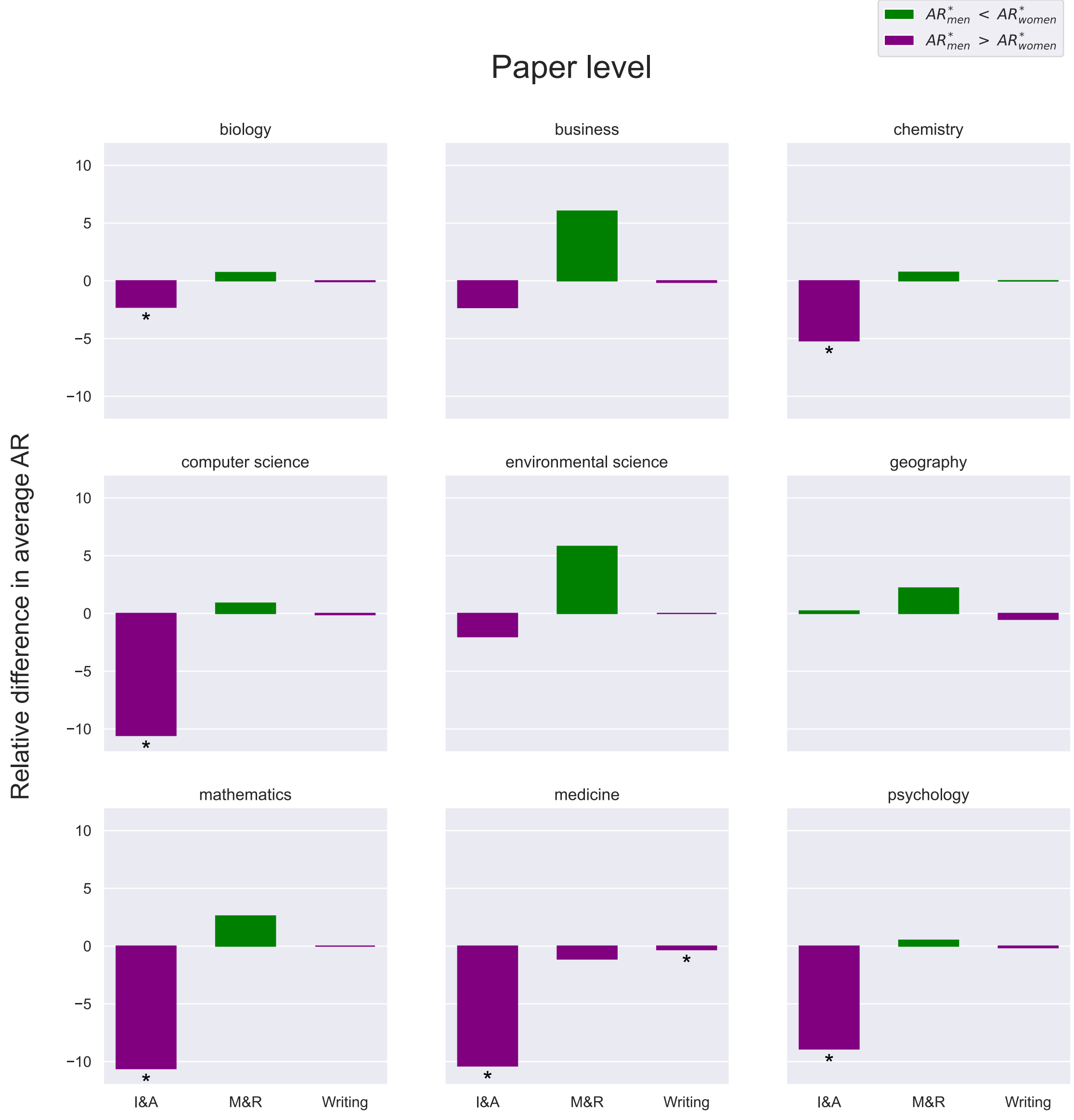

Fig. C1 Gender disparities in authorship rate across disciplines at the paper level. The relative difference in authorship rate ($AR$) between women and men is calculated using the same method as in Fig. 3. Asterisks (*) denote statistically significant differences ($p \leq 0.05$).

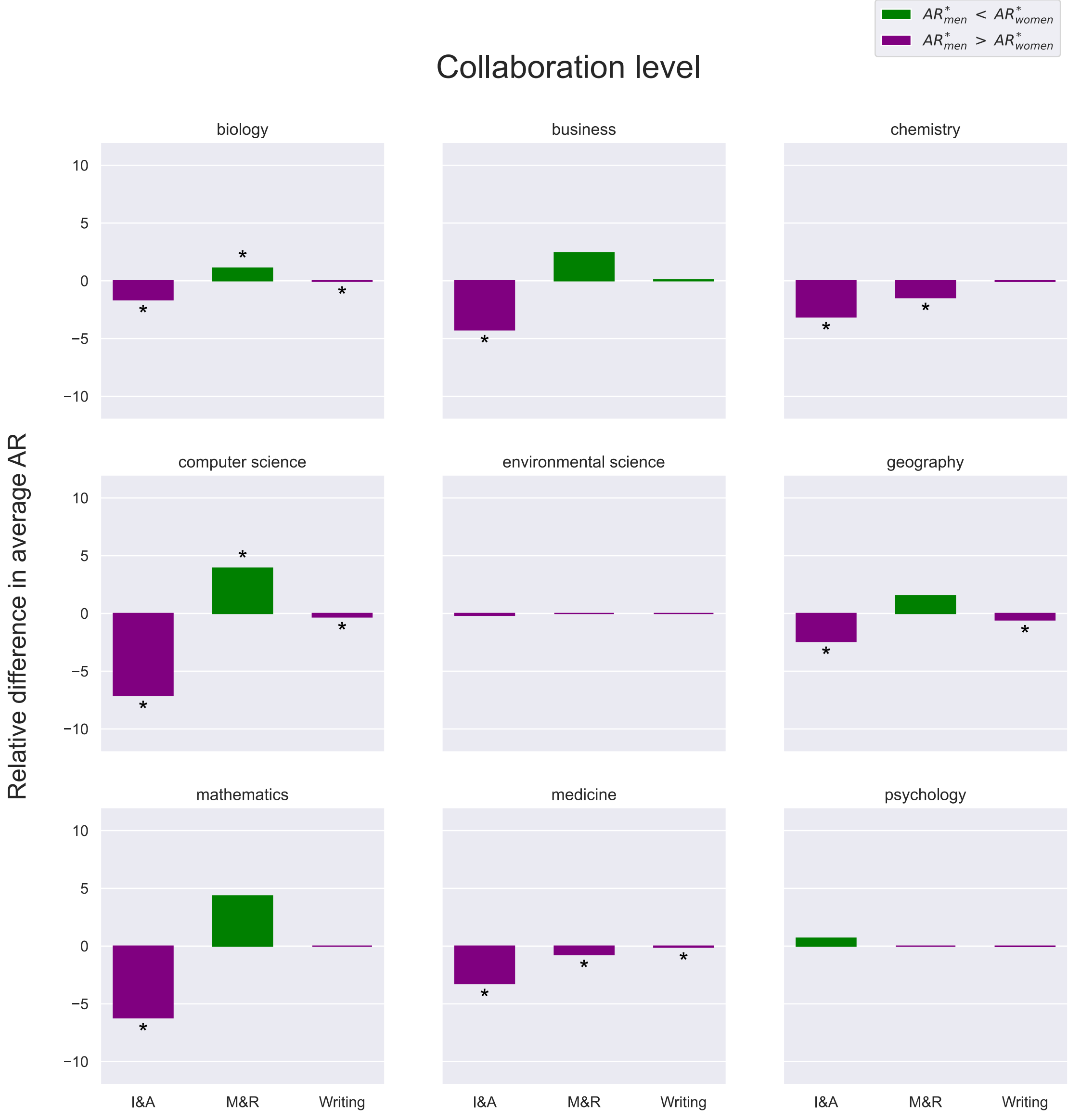

Fig. C2 Gender disparities in authorship rate across disciplines at the collaboration level using the same method as in Fig. C1.